\documentclass[submission,copyright,creativecommons]{eptcs}
\providecommand{\event}{QAPL 2013} % Name of the event you are submitting to\usepackage{etex}

\usepackage{aliascnt} %% this will make autoref give correct names
\def\mynewtheorem#1[#2]#3{%
  \newaliascnt{#1}{#2}%
  \newtheorem{#1}[#1]{#3}%
  \aliascntresetthe{#1}%
  \expandafter\def\csname #1autorefname\endcsname{#3}%
}
\usepackage{amsthm}

\newtheorem{theorem}{Theorem}

\mynewtheorem{lemma}[theorem]{Lemma}
\mynewtheorem{corollary}[theorem]{Corollary}
\mynewtheorem{proposition}[theorem]{Proposition}
\theoremstyle{definition}
\mynewtheorem{remark}[theorem]{Remark}
\mynewtheorem{convention}[theorem]{Convention}
\mynewtheorem{fact}[theorem]{Fact}
\mynewtheorem{example}[theorem]{Example}
\mynewtheorem{definition}[theorem]{Definition}
\mynewtheorem{property}[theorem]{Property}
\mynewtheorem{notation}[theorem]{Notation}

\usepackage{subcaption}
\usepackage{pdfpages}

\usepackage{graphics}
\usepackage[latin1]{inputenc} 
\usepackage[english]{babel} 
\usepackage{amsmath} 
\usepackage{amsfonts} 
\usepackage{amssymb} 
\usepackage{xspace} 
\usepackage{latexsym} 
\usepackage{url} 
\usepackage{xspace} 
\usepackage{graphics,color}              
\usepackage{array}
\usepackage{listings}
\lstdefinelanguage{IMP}
    {morekeywords={while,do,skip,goto,if,then,else},
     keywordstyle=\sffamily,
     moredelim=**[is][\color{blue}]{|}{|},
     mathescape,
     literate=*{'a}{$\alpha$}1 {'b}{$\beta$}1 {'c}{$\gamma$}1,
     basicstyle=\small,
    }
\usepackage{multirow}
\newcolumntype{b}{@{}>{{}}}
\newcolumntype{B}{@{}>{{}}c<{{}}@{}}
\newcolumntype{h}[1]{@{\hspace{#1}}}
\newcolumntype{L}{>{$}l<{$}}
\newcolumntype{C}{>{$}c<{$}}
\newcolumntype{R}{>{$}r<{$}}
\newcolumntype{S}{>{$(}r<{)$}}
\newcolumntype{n}{@{}}

\usepackage[%
    disable,
    colorinlistoftodos]{todonotes}
\usepackage{enumerate}
\usepackage{tikz}

\newcommand{\hbra}{\noindent\hbox to \textwidth{\leaders\hrule height1.8mm
depth-1.5mm\hfill}} 
\newcommand{\hket}{\noindent\hbox to \textwidth{\leaders\hrule
height0.3mm\hfill}} 
\newcommand{\ratio}{.3}

\newcommand{\Figbar}{{\center \rule{\hsize}{0.3mm}}}
\newcommand{\la}{\langle}               % the brackets for pairing (see also \pair) 
\newcommand{\ra}{\rangle}

\newcommand{\card}{\sharp}              % cardinality 
\newcommand{\s}[1]{{\sf #1}}    % sans-serif  

\newcommand{\eval}{\Downarrow}

\newcommand{\imp}{{\sf Imp}}            %imp language
\newcommand{\tern}[3]{#1\mathrel ? #2 : #3}
\newcommand{\sop}[1]{\s{#1}\ }
\newcommand{\sbin}[1]{\ \s{#1}\ }
\newcommand{\Ell}{\mathcal L}
\newcommand{\alphab}{A}
\newcommand{\betab}{B}
\newcommand{\gramm}{\mathrel{::=}}
\newcommand{\ass}{\mathrel{:=}}
\newcommand{\setof}[1]{\{\,#1\,\}}
\renewcommand{\to}[1][]{\stackrel{#1}{\rightarrow}}

\newcommand{\eg}{\emph{e.g.\ }}
\newcommand{\ie}{\emph{i.e.\ }}

\usetikzlibrary{decorations.pathreplacing}
\usetikzlibrary{calc}
\usetikzlibrary{positioning}

\def\lbl#1:{\mbox{\color{blue}$#1$}:}
\def\Lbl#1<#2>:{\mbox{\color{blue}$#1\langle #2\rangle$}:}

\tikzset{
    every picture/.style={
        baseline={([yshift=-.5ex]current bounding box)},
        remember picture,
    },
    lstnode/.style={inner ysep=-4pt,inner xsep=3pt,outer sep=3pt,draw,rounded corners=5pt},
}
\lstnewenvironment{lstnode}[3][]
{
 \node[lstnode,#1] (#2)
 \bgroup
 \minipage{#3}
}
{\endminipage\egroup;}

\let\oldparagraph\paragraph
\makeatletter
\def\paragraph{\@ifnextchar*\new@paragraph@star\new@paragraph}
\def\new@paragraph@star*#1{\oldparagraph*{#1.}}
\def\new@paragraph#1{\oldparagraph{#1.}}
\makeatother

\title{Indexed Labels for Loop Iteration Dependent Costs}
\def\titlerunning{Indexed Labels for Loop Iteration Dependent Costs}
\author{Paolo Tranquilli\footnote{This work is funded by the CerCo FET-Open EU Project.}%
    \institute{DISI (Dipartimento di Informatica -- Scienza e Ingegneria)\\Universit\`a di Bologna Alma Mater}%
    \email{tranquil@cs.unibo.it}%
}
\def\authorrunning{Paolo Tranquilli}

\begin{document}

\maketitle

\begin{abstract}
We present an extension to the labelling approach, a technique for lifting
resource consumption information from compiled to source code. This approach,
which is at the core of the annotating compiler from a large fragment of C to
8051 assembly of the CerCo project, loses preciseness when differences arise
as to the cost of the same portion of code, whether due to code transformation
such as loop optimisations or advanced architecture features (\eg cache). We
propose to address this weakness by formally indexing cost labels with the
iterations of the containing loops they occur in. These indexes can be
transformed during the compilation, and when lifted back to source code they
produce dependent costs.

The proposed changes have been implemented in CerCo's untrusted prototype
compiler from a large fragment of C to 8051 assembly.
\end{abstract}
\listoftodos

\section{Introduction}
\label{sec:intro}
Recent years have seen impressive advancements in the field of formal description
and certification of software components. In the fields of compilers a well-documented
example is CompCert, a project which has spawned the proof of correctness of a
compiler from a large fragment of C to assembly~\cite{CompCert}. The success
of this endeavour is also supported by a comparison with other compilers as
to the number of bugs found with testing tools~\cite{findingbugs}.

The CerCo project~\cite{cerco} strives to add a significant aspect to the
picture: certified resource consumption. More precisely our aim is to build
a certified C compiler targeting embedded systems that produces, apart
from object code functionally equivalent to the input, an \emph{annotation} of the
source code which is a sound and precise
description of the execution cost of the compiled code. Time and stack are the
immediate resources on which the method can be applied.

The current state of the art in commercial products that analyse
reaction time or memory usage of programs installed in embedded systems
(\eg Scade~\cite{scade} or AbsInt~\cite{absint}) is that the estimate is based
upon an abstract interpretation of the object code that may require explicit and
untrusted annotations of the binaries stating how many times loops are iterated
(see \eg \cite{WCETsurvey}). Our aim, on the other hand, is to lift cost information
of small fragments of object code, so that these bits of
information may be compositionally combined at the source level,
abstracting away the specifics of the architecture and only having to reason
about standard C semantics the programmer will be familiar with. This
information can be used to decide complexity
assertions either with pencil and paper or
with a tool for automated and formal reasoning about C programs such as
Frama-C~\cite{framac}.

The theoretical basis of the CerCo compiler has been outlined by Amadio \emph{et al}~\cite{labeling},
where in particular the labelling approach is described.
Summarising, the proposal consists in `decorating' the source code by inserting
labels at key points.
These labels are preserved as compilation progresses, from one intermediate language to another.
Once the final object code is produced, such labels should correspond to the parts of the compiled code that have a constant cost. This cost can then be assigned to blocks of source code.

Two properties must hold of any cost estimate given to blocks of code.
The first property, paramount to the correctness of the method, is \emph{soundness}---the actual execution cost must be bounded by the estimate.
In the labelling approach, this is guaranteed if every loop in the control flow of the compiled code passes through at least one cost label. Were it not the case, the cost of the loop would be taken in charge by a label external to it, so that any constant cost assignment
would be invalidated by enough iterations of the loop.
The second property, optional but desirable, is \emph{preciseness}---the estimate
\emph{is} the actual cost. This is of particular importance for embedded real-time
systems, where in particular situations we may care that a code runs for \emph{at least}
some clock cycles.
In the labelling approach, this is true if, for every label, every possible execution of the compiled code starting from such a label yields the same cost before hitting another one.
In simple architectures such as the 8051 micro-controller which is targeted by the
current stage of the CerCo project, this can be guaranteed by placing labels at the
start of any branch in the control flow, and by ensuring that no labels are duplicated.

The reader should note that the above mentioned requirements state properties
that must hold for the code obtained \emph{at the end} of the compilation chain.
Even if one is careful about injecting the labels at suitable places in the source
code, the requirements might still fail because of two main obstacles.
\begin{itemize}
\item
The compilation process might introduce important changes in the control flow, inserting loops or branches.
This might happen for example when replacing operations that are unavailable in the target architecture, such as
generic shift and multi-byte division in the 8051 architecture%
\footnote{The reader might see the work outlined in \cite{D2.2} to get a grasp of how we tackle this problem in CerCo's compiler.}.
\item
Even when the compiled code \emph{does}---as far as the the syntactic control flow graph is concerned---respect the conditions for soundness and preciseness, the cost of blocks of instructions might not be
independent of context and thus not compositional, so that different passes through a label might have different costs.
This becomes a concern if one wishes to apply the approach to more complex architectures, for example one with caching or pipelining.
\end{itemize}
Even if we solved the problem outlined in the first point for our current compilation chain, the point remains
a weakness of the current labelling approach when it comes to some common code transformations.
In particular, most \emph{loop optimisations} change the control flow graph duplicating code and adding or changing the branches.
An example optimisation of this kind is \emph{loop peeling}, where a first iteration of the loop is hoisted out of and before its body.
This optimisation is employed by compilers in order to trigger other optimisations, such as dead code elimination or invariant code motion.
Here, the hoisted iteration might possibly be assigned a different cost than later iterations.

The second point above highlights another weakness. Different tools allow to predict up to a certain extent the behaviour of cache.
For example, the \s{aiT} tool~\cite{absint} allows the user to estimate the worst-case execution time taking into account advanced features of the target architecture. While
such a tool is not fit for a compositional approach which is central to CerCo's project\footnote{\s{aiT} assumes the cache is empty at the start of computation, and treats each procedure call separately, unrolling a great part of the control flow.},
\s{aiT}'s ability to produce tight estimates of execution costs would still enhance the effectiveness of the CerCo compiler, \eg{} by integrating such techniques in its development.
A typical case where cache analysis yields a difference in the execution cost of a block is in loops: the first iteration will usually stumble upon more cache misses than subsequent iterations.

If one looks closely, the source of the two weaknesses of the regular labelling approach of~\cite{labeling} outlined above is common: the inability to state different costs for different occurrences of labels in the execution trace. The difference in cost might be originated by labels being duplicated along the compilation, or by the costs being sensitive to the current state of execution.

The work we present here addresses this weakness by introducing cost labels that are dependent on which iteration of its containing loops it occurs in.
This is achieved by means of \emph{indexed labels}; all cost labels are decorated with formal indexes coming from the loops containing such labels.
These indexes allow us to rebuild, even after multiple loop transformations, which iterations of the original loops in the source code a particular label occurrence belongs to.
During the annotating stage, this information is presented to the user by means of \emph{dependent costs}.

Here we concentrate on integrating the labelling approach with two loop transformations---\emph{loop peeling} and \emph{loop unrolling}.
They will be presented for a toy language in \autoref{sec:defimp},
For general information on compiler optimisations (and loop optimisations in particular) we refer the reader to the vast literature on the subject (\eg\cite{muchnick,morgan}).

The proposed changes have been implemented in CerCo's untrusted prototype compiler
available on CerCo's homepage\footnote{\url{http://cerco.cs.unibo.it/}}. For
lack of space the present work will not delve into the details of the implementation.

Whilst we cover only two loop optimisations in this paper, we argue that the work presented herein poses a good foundation for extending the labelling approach, in order to cover more and more common optimisations, as well as gaining insight into how to integrate advanced cost estimation techniques, such as cache analysis, into the CerCo compiler.
Moreover loop peeling itself has the fortuitous property of enhancing and enabling other optimisations.
Experimentation with CerCo's untrusted prototype compiler, which implements constant propagation and partial redundancy elimination~\cite{PRE,muchnick}, show how loop peeling enhances those other optimisations.

\paragraph*{Outline}
We will present our approach on a minimal `toy' imperative language, \imp{} with \s{goto}s, which we present in \autoref{sec:defimp} along with formal definitions of the loop transformations.
This language already presents most of the difficulties encountered when dealing with C,
so we stick to it for the sake of this presentation.
In \autoref{sec:labelling} we summarize the labelling approach as presented in~\cite{labeling}.
\hyperref[sec:indexedlabels]{Section~\ref*{sec:indexedlabels}} presents \emph{indexed labels}, our proposal for dependent labels which are able to describe precise costs even in the presence of the various loop transformations we consider, together with a more detailed example
(\autoref{ssec:detailedex}).
Finally \autoref{sec:future} speculates on further work on the subject.

\section{The minimal imperative language \imp{}}\label{sec:defimp}
We briefly outline the toy language, the minimalist imperative language \imp{}.
Its syntax
% and (small-step) operational semantics are
is presented in~\autoref{fig:minimp}.
\begin{figure}
$$\begin{gathered}
\begin{array}{nlBl>(R<)n}
% \multicolumn{4}{C}{\bfseries Syntax}\\
\multicolumn{4}{ncn}{
%   \ell,\ldots \hfill \text{(labels)}\hfill
x,y,\ldots \hfill
\text{(identifiers)}
\hfill e,f,\ldots \hfill \text{(expressions)}
}\\
P,S,T,\ldots &\gramm& \s{skip} \mid s;t
\mid \sop{if}e\sbin{then}S\sbin{else}T
\mid \sop{while} e \sbin{do} s \mid x \ass e & statements\\
% \\
% \multicolumn{4}{C}{\bfseries Semantics}\\
% K,\ldots  &\gramm& \s{halt} \mid S \cdot K & continuations
% \end{array}
% \\[15pt]
% % \s{find}(\ell,S,K) \ass
% % \left\{\begin{array}{lL}
% % \bot & if $S=\s{skip},\sop{goto} \ell'$ or $x\ass e$,\\
% % (T, K) & if $S=\ell:T$,\\
% % \s{find}(\ell,T,K) & otherwise, if $S = \ell':T$,\\
% % \s{find}(\ell,T_1,T_2\cdot K) & if defined and $S=T_1;T_2$,\\
% % \s{find}(\ell,T_1,K) & if defined and
% % $S=\sop{if}b\sbin{then}T_1\sbin{else}T_2$,\\
% % \s{find}(\ell,T_2,K) & otherwise, if $S=T_1;T_2$ or
% % $\sop{if}b\sbin{then}T_1\sbin{else}T_2$,\\
% % \s{find}(\ell,T,S\cdot K) & if $S = \sop{while}b\sbin{do}T$.
% % \end{array}\right.
% % \\[15pt]
% \begin{array}{lBl}
% (x:=e,K,s)  &\to& (\s{skip},K,s[v/x]) \qquad\mbox{if }(e,s)\eval v \\ \\
% 
% (S;T,K,s)  &\to& (S,T\cdot K,s) \\ \\
% 
% (\s{if} \ b \ \s{then} \ S \ \s{else} \ T,K,s)
% &\to&\left\{
% \begin{array}{ll}
% (S,K,s) &\mbox{if }(b,s)\eval v \neq 0 \\
% (T,K,s) &\mbox{if }(b,s)\eval 0
% \end{array}
% \right. \\ \\
% 
% 
% (\s{while} \ b \ \s{do} \ S ,K,s)
% &\to&\left\{
% \begin{array}{ll}
% (S,\s{while} \ b \ \s{do} \ S \cdot K,s) &\mbox{if }(b,s)\eval v \neq 0 \\
% (\s{skip},K,s) &\mbox{if }(b,s)\eval 0
% \end{array}
% \right. \\ \\
% 
% 
% (\s{skip},S\cdot K,s)  &\to&(S,K,s) % \\ \\
% 
% % (\ell : S, K, s)  &\to& (S,K,s) \\ \\
% 
% % (\sop{goto}\ell,K,s)  &\to& (\s{find}(\ell,P,\s{halt}),s) \\ \\
\end{array}
\end{gathered}$$
\caption{The syntax
% and operational semantics
of \imp.}
\label{fig:minimp}
\end{figure}
We may omit the \s{else} clause of a conditional if it leads to a \s{skip} statement.
The precise grammar for expressions is not particularly relevant so we do not give one in full.
We will use the notation
$(S, K, s) \to (S', K', s')$ for \imp's small-step semantics
of which we skip the unsurprising definition.
$S$ is the statement being executed,
$K$ is a continuation (\ie a stack of statements to be executed after $S$)
and $s$ is the store (\ie a map from variables to integers).
%  The statement $(e,s)\Downarrow v$
% means that $e$ evaluates to value $v$ in the environment $s$ (which is
% a map from identifiers to integers). The update of the environment is written
% with $s[v/x]$---assigning $v$ to $x$ in $s$.

% We will presuppose that all programs are \emph{well-labelled}, \ie every label labels at most one occurrence of a statement in a program, and every \s{goto} points to a label actually present in the program.
% The \s{find} helper function has the task of not only finding the labelled statement in the program, but also building the correct continuation.
% The continuation built by \s{find} replaces the current continuation in the case of a jump.

\paragraph*{Further down the compilation chain}
We abstract over the rest of the compilation chain.
We posit the existence, for every language $L$ further down the compilation chain, of a suitable notion of `sequential instructions', wherein each instruction has a single natural successor. To these sequential instructions we can add our own.

\begin{figure}
 \centering
\begin{tabular}{n*{5}{cn}}
% Loop peeling&&&&Loop unrolling\\
\begin{tikzpicture}[lstnode]
\begin{lstnode}{peel2}{3.2cm}
if $b$ then
    $S$;
    while $b$ do $S$
\end{lstnode}
\end{tikzpicture}
&
\tikz[baseline={([yshift=-.5ex]n.mid)}]\node(n)[inner sep=0,rotate=180]{${}\mapsto{}$};
&
\begin{tikzpicture}[lstnode]
\begin{lstnode}{peel1}{2.3cm}
while $b$ do $S$
\end{lstnode}
\end{tikzpicture}
&
${}\mapsto{}$
% where $\ell'_i$ is a fresh label for any $\ell_i$ labelling a statement in $S$.
% This relabelling is safe for \s{goto}s occurring outside the loop because of the single-entry condition.
% Note that for \s{break} and \s{continue} statements, those should be replaced with \s{goto}s in the peeled body $S$.
&
\begin{tikzpicture}[lstnode]
\begin{lstnode}{lblunroll2}{3.2cm}
while $b$ do
    $S$;
    if $b$ then
        $S$;
        $\vdots$
        if $b$ then
            $S$
\end{lstnode}
\end{tikzpicture}
\end{tabular}
\caption{Loop peeling (left) and loop unrolling (right).}
\label{fig:loop_transformations}
\end{figure}
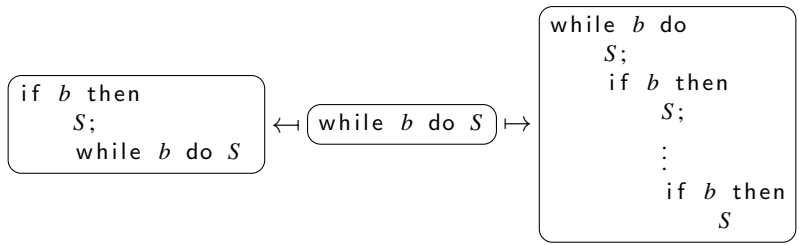

\paragraph*{Loop transformations}
% We call a loop $L$ \emph{single-entry} in $P$ if there is no \s{goto} to $P$ outside of $L$ which jumps into $L$.\footnote{This is a reasonable aproximation: it defines a loop as multi-entry if it has an external but unreachable \s{goto} jumping into it.}
% Many loop optimisations do not preserve the semantics of multi-entry loops in general, or are otherwise rendered ineffective.
% Usually compilers implement a single-entry loop detection which avoids the multi-entry ones from being targeted by optimisations~\cite{muchnick,morgan}.
We present the loop transformations we deal with in \autoref{fig:loop_transformations}.
These transformations are local, \ie they target a single loop and transform it.
Which loops are targeted may be decided by some \emph{ad hoc} heuristic.
However, the precise details of which loops are targeted and how is not important here.

As already mentioned in the introduction, loop peeling consists in preceding the loop with a copy of its body, appropriately guarded.
This is usually done to trigger further optimisations.
% , such as those that rely on execution information which can be computed at compile time, but which is erased by further iterations of the loop, or those that use the hoisted code to be more effective at eliminating redundant code.
Integrating this transformation into the labelling approach would also allow, in the future, the integration of a common case of cache analysis, as predicting cache hits and misses benefits from a form of \emph{virtual} loop peeling~\cite{cacheprediction}.

Loop unrolling consists of the repetition of several copies of the body of the loop inside the loop itself (inserting appropriate guards, or avoiding them altogether if enough information about the loop's guard is available at compile time).
This can limit the number of (conditional or unconditional) jumps executed by the code and trigger further optimisations dealing with pipelining, if appropriate for the architecture.
Notice that we present unrolling in a wilfully \emph{na\"{i}ve} version.
On the one hand usually
less general loops and more well-behaving loops are targeted; on the other hand,
conditionals are seldom used to cut up the body of the unrolled loop.
However we are mainly interested in the changes to the control flow the transformation
does. The problem this transformation poses to CerCo's labelling approach are
independent of the sophistication of the actual transformation.

We decided to apply transformations in the front-end in order to only target loops explicitly
written by the programmer. This is because we need to output source code annotations that are
meaningful to the user, and in order to do so we only transform loops that were explicitly
written as so.

% where $\ell^j_i$ are again fresh labels for any $\ell_i$ labelling a statement in $S$.

\begin{example}
In \autoref{fig:example1} we show a program (a wilfully inefficient computation of the
sum of the first $n$ factorials) and a possible combination of transformations
applied to it (again for the sake of presentation rather than efficiency).
\begin{figure}[!ht]
% \lstset{numbers=left,numberstyle=\tiny, stepnumber=1,}
% \centering
\begin{tikzpicture}
\begin{lstnode}{code1}{3.21cm}
$s\ass 0$;
$i\ass 0$;
while $i<n$ do
    $p\ass 1$;
    $j\ass 1$;
    while $j\le i$ do
        $p\ass j*p$;
        $j\ass j+1$;
    $s\ass s+p$;
    $i\ass i+1$;
\end{lstnode}
\begin{lstnode}[right=1em of code1]{code2}{6cm}
$s\ass 0$;
$i\ass 0$;
if $i<n$ then
    $p\ass 1$;
    $j\ass 1$;
    while $j\le i$ do
        $p\ass j*p$;
        $j\ass j+1$;
    $s\ass s+p$;
    $i\ass i+1$;
    while $i<n$ do
        $p\ass 1$;
        $j\ass 1$;
        if $j\le i$ then
            $p\ass j*p$;
            $j\ass j+1$;
            if $j\le i$ then
                $p\ass j*p$;
                $j\ass j+1$;
                while $j\le i$ do
                    $p\ass j*p$;
                    $j\ass j+1$;
                    if $j\le i$ then
                        $j\ass j+1$;
        $s\ass s+p$;
        $i\ass i+1$;
        if $i<n$ then
            $p\ass 1$;
            $j\ass 1$;
            while $j\le i$ do
                $p\ass j*p$;
                $j\ass j+1$;
                if $j\le i$ do
                    $p\ass j*p$;
                    $j\ass j+1$;
            $s\ass s+p$;
            $i\ass i+1$;
\end{lstnode}
\draw [thick,|->] (code1) -- (code2);
\coordinate (a) at ([shift={(8pt,-5pt)}]code2.north west);
\coordinate (b) at ([shift={(8pt,7.5pt)}]code2.south west);
% \foreach \i in {0,...,38}
%     \draw[blue] ($(a)!\i/38!(b)$) coordinate (tmp) -- (tmp-|code2.west);
\foreach \x/\y/\s/\t in 
    {19.625/24.75/0/unrolled,13.325/24.75/1/peeled,29.875/36/0/unrolled,
     10.25/38/2/unrolled,2/38/3/peeled} {
     \foreach \a/\b in {\x/a,\y/b}
        \draw [ultra thin, gray, densely dashed]
            ($(a)!\a/38!(b)$) coordinate (tmp_\b) --
            (tmp_\b-|code2.east) -- ++($\s*(17.5pt,0)$) coordinate (tmp_\b);
     \draw[solid,thick,decorate, decoration={brace, amplitude = 7.5pt}]
         (tmp_a) -- node[sloped, anchor = base, yshift = 9pt]{\t} (tmp_b);
}
\end{tikzpicture}
\caption{An example of loop transformations. Blocks are delimited by indentation.}
\label{fig:example1}
\end{figure}
\end{example}

\section{Labelling: a quick sketch of the previous approach}
\label{sec:labelling}
Plainly labelled $\ell$\imp{} is obtained by adding to the code \emph{cost labels}
(with metavariables $\alpha,\beta,\ldots$), and cost-labelled statements:
$$
S,T\gramm \cdots \mid \lbl\alpha: S
$$
Cost labels allow us to track some program points along the compilation chain.
For further details we refer to~\cite{labeling}.

The small step semantics turns into a labelled transition system
and a natural notion of trace (\ie lists of labels) arises. The small-step rules
of \imp{} remain as unlabelled steps, while adding the rule
$$(\lbl\alpha: S,K,s) \to[\alpha] (S,K,s)$$
% Here, we identify cost labels $\alpha$ with singleton traces and we use $\varepsilon$ for the empty trace.
Cost labels are thus emitted by cost-labelled statements only\footnote{In the general case,
because of the conditional ternary operator, any evaluation of expressions can emit cost labels too.}.
We then write $\to[\lambda]\!\!^*$ for the transitive closure of the small step semantics which produces by concatenation the trace $\lambda$.

\paragraph*{Labelling}
Given an \imp{} program $P$ its \emph{labelling}
in $\ell$\imp{} is defined by $\lbl\alpha:\Ell(P)$,
putting cost labels after every branching statement, at the start of both branches, and a cost label at the beginning of the program.
% Also, every labelled statement gets a cost label, which is a conservative approach to ensuring that all loops have labels inside them, as a loop might be done with \s{goto}s.
The relevant recursive cases for the definition of $\Ell(P)$ are
$$\begin{aligned}
  \Ell(\sop{if}e\sbin{then}S\sbin{else}T) &=
    \sop{if}e\sbin{then}\lbl\alpha:\Ell(S)\sbin{else}\lbl\beta:\Ell(T)\\
  \Ell(\sop{while}e\sbin{do}S) &=
    (\sop{while}e\sbin{do}\lbl\alpha:\Ell(S));\lbl\beta:\s{skip}\\
%   \Ell(\ell : S) &=
%     (\ell : \alpha : \Ell(S))
  \end{aligned}$$
where $\alpha,\beta$ are fresh cost labels.
In all other cases the definition just passes to substatements. Notice that
labelling enjoys soundness (a label is added inside each loop) and preciseness
(there is a label at all branches, included the loop-exiting one).

\paragraph*{Labels in the rest of the compilation chain}
All languages further down the chain get a new sequential statement $\sop{emit}\alpha$ whose effect is to be consumed in a labelled transition while keeping the same state.
All other instructions guard their operational semantics and do not emit cost labels.

Preservation of semantics throughout the compilation process is restated, in rough terms, as:
\begin{equation}\label{eq:preservation}
\text{starting state of $P$}\to[\lambda]\!\!^*\;\text{halting state} \iff
\text{starting state of $\mathcal C(P)$} \to[\lambda]\!\!^*\;\text{halting state}
\end{equation}
Here $P$ is a program of a language along the compilation chain, starting and halting states depend on the language, and $\mathcal C$ is any of the compilation passes\footnote{The case of divergent computations needs to be addressed too.
Also, the requirement can be weakened by demanding a weaker form of equivalence of the traces than equality.
Both of these issues are beyond the scope of this presentation.}. This must in
particular be true for any optimisation pass the compilation undergoes.

\paragraph*{Instrumentations}
Let $\mathcal C$ be the whole compilation from $\ell\imp$ to the labelled version of some low-level language $L$.
Supposing such compilation has not introduced any new loop or branching, we have that:
\begin{itemize}
\item
every loop contains at least a cost label;
\item
every branching has different labels for the two branches.
\end{itemize}
With these two conditions, we have that each and every cost label in $\mathcal C(P)$ for any $P$ corresponds to a block of sequential instructions, to which we can assign a constant \emph{cost}\footnote{This in fact requires the machine architecture to be `simple enough', or for some form of execution analysis to take place.}.
As we have explained in the \hyperref[sec:intro]{introduction}, the two
properties above ensure \emph{soundness} and \emph{preciseness} of this
cost estimate respectively.
We therefore may assume the existence of a \emph{cost mapping} $\kappa_P$ from cost labels to natural numbers, assigning to each cost label $\alpha$ the cost of the block containing the single occurrence of $\alpha$.

Given any cost mapping $\kappa$, we can enrich a labelled program so that a particular fresh variable (the \emph{cost variable} $c$) keeps track of the summation of costs during the execution.
We call this procedure \emph{instrumentation} of the program, and it is defined recursively by:
$$
\mathcal I(\lbl\alpha:S) = c \ass c + \kappa(\alpha) ; \mathcal I(S)
$$
In all other cases the definition passes to substatements. One can then reason
on the instrumented version of the code like he would on any program, asserting
statements about complexity by inspecting $c$.

\paragraph*{The problem with loop optimisations}
Let us take loop peeling, and apply it to the labelling of a program without any prior adjustment:
$$
(\sop{while}e\sbin{do}\lbl\alpha:S);\lbl\beta:\s{skip}
\mapsto
(\sop{if}b\sbin{then} \lbl\alpha: S; \sop{while} b \sbin{do} \lbl\alpha:
S);
\lbl\beta:\s{skip}
$$
What happens is that the cost label $\alpha$ is duplicated with two distinct occurrences.
If these two occurrences correspond to different costs in the compiled code, the best the cost mapping can do is to take the maximum of the two, preserving soundness (\ie the cost estimate still bounds the actual one) but losing preciseness (\ie the actual cost could be strictly less than its estimate).

\section{Indexed labels}
\label{sec:indexedlabels}
This section presents the core of the new approach.
In brief points it amounts to the following:
\edef\ssec{ssec}
\label{ssec7}%just to silence stupid warning
\begin{enumerate}%
[\bfseries\expandafter\ref\expandafter{\expandafter \ssec\arabic{enumi}}.]
\item
Enrich cost labels with formal indexes stating, for each loop containing the label
in the source code, what iteration it occurs in.
\item
Each time a loop transformation is applied and a cost labels is split in different occurrences, each of these will be reindexed so that every time they are emitted their position in the original loop will be reconstructed.
\item
Along the compilation chain, alongside the \s{emit} instruction we add other instructions updating the indexes, so that iterations of the original loops can be rebuilt at the operational semantics level
even when the original structure of loops is lost.
\item
The machinery computing the cost mapping will still work, but assigning costs to indexed cost labels, rather than to cost labels as we wish.
However, \emph{dependent costs} can be calculated, where dependency is on which iteration of the containing loops we are in.
\end{enumerate}

\subsection{Indexing the cost labels}
\label{ssec:indlabs}
\label{ssec1}

\paragraph*{Formal indexes and $\iota\ell\imp$}
Let $i_0,i_1,\ldots$ be a sequence of distinguished fresh identifiers that will be used as loop indexes.
A \emph{simple expression} is an affine arithmetical expression in one of these indexes, that is $a*i_k+b$ with $a,b,k \in \mathbb N$.
\label{pag:exprcomp}Simple expressions $e_1=a_1*i_k+b_1$ and $e_2=a_2*i_k+b_2$ in the same index can be composed---substituting $e_2$ in the $i_k$ of $e_1$ we have $e_1\circ e_2\ass (a_1a_2)*i_k + (a_1b_2+b_1)$, and this operation has an identity element $1*i_k+0$ (which we will denote simply by $i_k$).
Constants can be expressed as simple expressions, so that we identify a natural $c$ with $0*i_k+c$.

An \emph{indexing} (with metavariables $I$, $J$, \ldots) is a list of transformations of successive formal indexes dictated by simple expressions, that is a mapping%
    \footnote{Here we restrict each mapping to be one from an index to a
    simple expression \emph{on the same index}. This might not be the case if more loop
    optimisations are accounted for (for example, interchanging two nested
    loops could give rise to an indexing like $i_0\mapsto i_1,i_1\mapsto i_0$).}
$$
i_0\mapsto a_0*i_0+b_0,\dots, i_{k-1} \mapsto a_{k-1}*i_{k-1}+b_{k-1}
$$

An \emph{indexed cost label} (metavariables $\alphab$, $\betab$, \ldots) is the combination of a cost label $\alpha$ and an indexing $I$, written $\alpha\la I\ra$.
The cost label underlying an indexed one is called its \emph{atom}.
% All plain labels can be considered as indexed ones by taking an empty indexing, \ie $\alpha\la\ra$.

\imp{} with indexed labels (from now on $\iota\ell\imp$) is defined by
having loops with a formal index attached to them and by allowing statements to
be labelled by indexed labels:
$$
S,T,\ldots \gramm \cdots \lbl{i_k}:\sop{while}e\sbin{do}S\mid \lbl\alphab : S
$$
Notice that unindexed loops may still exist in the language: though it does not
concern this simple toy example, they would correspond to multi-entry loops which
are ignored by indexing and optimisations in a scenario with gotos.

We will discuss $\iota\ell$\imp{}'s semantics later, in \autoref{ssec:inlabsem}.

\paragraph*{Indexed labelling}
% Given an $\imp$ program $P$, in order to index loops and assign indexed labels, we must first distinguish single-entry loops.
% We sketch how this can be computed in the sequel.
% 
% A first pass of the program $P$ can easily compute two maps: $\s{loopof}_P$ from each label $\ell$ to the occurrence (\ie the path) of a $\s{while}$ loop containing $\ell$, or the empty path if none exists; and $\s{gotosof}_P$ from a label $\ell$ to the occurrences of \s{goto}s pointing to it.
% Then the set $\s{multientry}_P$ of multi-entry loops of $P$ can be computed by
% $$
% \s{multientry}_P\ass\{\, p \mid \exists \ell,q.p =\s{loopof}_P(\ell),q\in\s{gotosof}_P(\ell), q \not\le p\,\}
% $$
% Here $\le$ is the prefix relation\footnote{Possible simplifications to this procedure include keeping track of just the while loops containing labels and \s{goto}s (rather than paths in the syntactic tree of the program), and making two passes while avoiding building the map to sets $\s{gotosof}$}.
In order to compute the \emph{indexed labelling} $\Ell^\iota$ of a program, we
need to keep track of the nesting of indexed loops as we visit the program
abstract syntax tree.

Let $Id_k$ be the indexing of length $k$ made from identity simple expressions,
\ie the sequence $i_0\mapsto i_0, \ldots , i_{k-1}\mapsto i_{k-1}$.
We define the tiered indexed labelling $\Ell^\iota (S,k)$ by recursion setting:
\begin{align*}
\Ell^\iota(\sop{while}b\sbin{do}T, k)
&\ass \lbl{i_k}:\sop{while}b\sbin{do}\lbl{\alpha\la Id_{k+1}\ra}: \Ell^\iota(T,k+1));\lbl{\beta\la Id_k \ra}: \s{skip}\\
\Ell^\iota(\sop{if}b\sbin{then}T_1\sbin{else}T_2, k)
&\ass \sop{if}b\sbin{then} \lbl{\alpha\la Id_k \ra} : \Ell^\iota(T_1,k) \sbin{else} \lbl{\beta\la Id_k \ra} : \Ell^\iota(T_2,k)
\end{align*}
Here, as usual, $\alpha$ and $\beta$ are fresh cost labels, and other cases just keep making the recursive calls on the substatements.
The \emph{indexed labelling} of a program $P$ is then defined as $\alpha\la \ra : \Ell^\iota(P,0)$, \ie a further fresh unindexed cost label is added at the start, and we start from level $0$.

In plainer words: each loop is indexed by $i_k$ where $k$ is the number of other loops containing this one, and all cost labels under the scope of a loop indexed by $i_k$ are indexed by all indexes $i_0,\ldots,i_k$, without any transformation.

\subsection{Indexed labels and loop transformations}\label{ssec2}
We define the \emph{reindexing} $\alpha\la I\ra\circ (i_k\mapsto f)$ as an operator on indexed labels by setting%
\footnote{%
    If mappings are not restricted to only depend on the index being mapped,
    reindexing should be substituted in each occurrence of $i_k$.}:
\begin{multline*}
\alpha \la i_0\mapsto e_0,\ldots, i_k \mapsto e_k,\ldots,i_n\mapsto e_n\ra
\circ(i_k\mapsto f)
\ass%\\
\alpha \la i_0\mapsto e_0,\ldots, i_k \mapsto e_k \circ f,\ldots,i_n\mapsto e_n \ra.
\end{multline*}
We extend this definition to statements in $\iota\ell\imp$
by applying the above transformation to all indexed labels contained in a statement.

\begin{figure}
\centering
\begin{tabular}{l@{${}\mapsto{}$}l}
% \multicolumn{2}{l}{Loop peeling:}\\
\begin{tikzpicture}[lstnode]
\begin{lstnode}{lblpeel1}{2.9cm}
$\lbl i_k:$ while $b$ do $S$
\end{lstnode}
\end{tikzpicture}
&
\begin{tikzpicture}[lstnode]
\begin{lstnode}[right=1em of lblpeel1]{lblpeel2}{8.5cm}
if $b$ then $S\circ (i_k\mapsto 0)$; $\lbl i_k:$ while $b$ do $S\circ(i_k\mapsto i_k + 1)$
\end{lstnode}
\end{tikzpicture}
\\[5pt]
% \multicolumn{2}{l}{Loop unrolling:}\\
\begin{tikzpicture}[lstnode]
\begin{lstnode}{lblunroll1}{2.9cm}
$\lbl i_k:$ while $b$ do $S$
\end{lstnode}
\end{tikzpicture}
&
\begin{tikzpicture}[lstnode]
\begin{lstnode}{lblunroll2}{5.3cm}
$\lbl i_k:$ while $b$ do
    $S\circ(i_k\mapsto n*i_k)$;
    if $b$ then
        $S\circ(i_k\mapsto n*i_k+1)$;
        $\vdots$
        if $b$ then
            $S\circ(i_k\mapsto n*i_k+n-1)$
\end{lstnode}
\end{tikzpicture}
\end{tabular}
\caption{Loop peeling and loop unrolling in the presence of indexed labels. In loop unrolling
$n$ is the number of times the loop is unrolled.}
\label{fig:indexed_loop_transformations}
\end{figure}
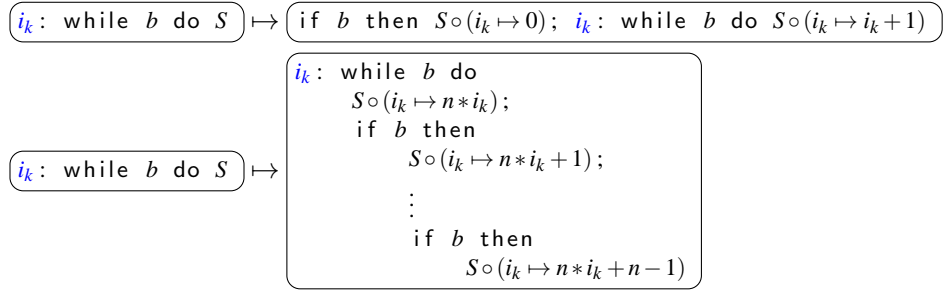

We can now finally redefine loop peeling and loop unrolling, taking into account indexed labels.
% It will only be possible to apply the transformation to indexed loops, that is loops that are single-entry.
The attentive reader will notice that no assumptions will be made as to the
labelling of the statements that are involved.
This ensures that the transformation can be repeated and composed at will.
Also, notice that after erasing all labelling information (\ie indexed cost labels and loop indexes) we recover exactly the same transformations presented in~\autoref{sec:defimp}. The transformations
are presented in \autoref{fig:indexed_loop_transformations}.

% where $n$ is the number of times the loop is unrolled.

As can be expected, in loop peeling the peeled iteration of the loop gets reindexed
with 0, as it always correspond to the first iteration of the loop.
The iterations of the remaining loop are shifted by $1$. Notice that this
transformation can lower the actual depth of some loops, however their index is
left untouched. In loop unrolling each copy of the unrolled body has its indexes
remapped so that when they are executed, the original iteration of the loop to which
they correspond can be recovered.

\begin{fact}
\label{fact:nonoverlap}
Loop peeling and unrolling preserve the following invariant, which we
call \emph{non-overlap of indexed labels}: for all
labels $\alpha\la I\ra$ and $\alpha\la J\ra$ such that $I\neq J$,
the first different simple expressions of the two are disjoint, \ie they always
evaluate to different constants.
Moreover for every loop
$\lbl{i_k}:\sop{while}e\sbin{do}S$ and label $\alpha\la I\ra$ in $S$,
no label outside the loop with the same atom can share the same prefix up to
$i_k$.
\end{fact}

\subsection{Semantics and compilation of indexed labels}\label{ssec:inlabsem}\label{ssec3}
In order to make sense of loop indexes, one must keep track of their values in the state.
A \emph{constant indexing} (metavariables $C,\ldots$) is an indexing which employs only constant simple expressions.
The evaluation of an indexed label $\alphab$ in a constant indexing $C$, denoted $\alphab|_C$, is defined by:
$$
\alphab|_{i_0\mapsto c_0,\ldots, i_{k-1}\mapsto c_{k-1}} \ass \alphab\circ(i_0\mapsto c_0)\circ\cdots\circ(i_{k-1}\mapsto c_{k-1})
$$
Here, we are using the definition of ${-}\circ{-}$ given in \autoref{ssec:indlabs} at \autopageref{pag:exprcomp}.
We consider the above defined only if the the resulting indexing of the label is constant too%
\footnote{%
    For example $(i_0\mapsto 2*i_0,i_1\mapsto i_1+1)|_{i_0\mapsto 2}$ is undefined, but $(i_0\mapsto 2*i_0,i_1\mapsto 0)|_{i_0\mapsto 2}= i_0\mapsto 4,i_1\mapsto 0$, is indeed a constant indexing, even if the domain of the original indexing is not covered by the constant one.%
}.
% The definition is extended to indexed labels by $\alpha\la I\ra|_C\ass \alpha\la I|_C\ra$.

Constant indexings will be used to keep track of the exact iterations of the original code that the emitted labels belong to.
We thus define two basic actions to update constant indexings: $C[i_k{\uparrow}]$ increments the value of $i_k$ by one, and $C[i_k{\downarrow}0]$ resets it to $0$.

We are ready to explain how the operational semantics of indexed labelled \imp{}
updates the one of plain $\ell\imp$.
The emitted cost labels will now be ones indexed by constant indexings.
We add to continuations a special indexed loop constructor
${i_k}:\sop{while} b \sbin {do} S \sbin{then} K$.

The difference between the regular stack concatenation $i_k:\sop{while}b\sbin{do}S\cdot K$ and the new constructor is that the latter indicates the loop is the active one in which we already are, while the former is a loop that still needs to be started\footnote{In the presence of \s{continue} and \s{break} statements active loops need to be kept track of in any case.}.
% The \s{find} function is updated accordingly with the case
% $$
% \s{find}(\ell, i_k:\sop{while}b\sbin{do}S, K) \ass \s{find}(\ell, S, i_k: \sop{while}b\sbin{do}S\sbin{then}K)
% $$

The state will now be a 4-tuple $(S,K,s,C)$ which adds a constant indexing to the triple of the regular semantics.
The small-step rules for all but cost-labelled and indexed loop statements remain the same,
without touching the $C$ parameter.
The new cases are:
$$\begin{aligned}
   (\lbl\alphab : S,K,s,C) &\to[\alphab|_C] (S,K,s,C)\\
   (\lbl{i_k}:\sop{while}b\sbin{do}S,K,C) &\to
    \begin{cases}
     (S,\lbl{i_k}:\sop{while}b\sbin{do}S\sbin{then} K,s,C[i_k{\downarrow}0])
     &\text{if $(b,s)\eval v\neq 0$,}\\
     (\s{skip}, K, s, C) &\text{otherwise,}
    \end{cases}\\
   (\s{skip}, \lbl{i_k}:\sop{while}b\sbin{do}S\sbin{then}K,C) &\to
    \begin{cases}
     (S,\lbl{i_k}:\sop{while}b\sbin{do}S\sbin{then} K,s,C[i_k{\uparrow}])
      & \text{if $(b,s)\eval v\neq 0$,}\\
     (\s{skip}, K, s, C) & \text{otherwise.}
    \end{cases}
  \end{aligned}$$
Here $(b,s)\Downarrow v$ means that expression $b$ evaluates to value $v$ in
memory state $s$. Some explanations are in order. We can see that
% \begin{itemize}
% \item
emitting a label always instantiates it with the current indexing, and that
% \item
hitting an indexed loop the first time initializes the corresponding index to 0. Continuing the same loop increments the index as expected.
% \item
% The \s{find} function ignores the current indexing: this is correct under the assumption that all indexed loops are single entry, so that when we land inside an indexed loop with a \s{goto}, we are sure that its current index is right.
% \item The value of a loop index is not forgotten upon exiting the loop. While
% it would be possible to do so in this setting, it is unneeded and becomes more
% complex in the presence of gotos. Once we exit a loop the last corresponding
% loop index will not contribute to reindexings.
% \item

The starting state with store $s$ for a program $P$ is $(P,\varepsilon,s,(i_0\mapsto 0,\dots,i_{n-1}\mapsto 0)$ where
$\varepsilon$ is the empty stack and $i_0,\ldots,i_{n-1}$ cover all loop indexes of $P$\footnote{For a program which is the indexed labelling of an \imp{} one this corresponds to the maximum nesting of single-entry loops.
We can also avoid computing this value in advance if we define $C[i{\downarrow}0]$ to extend $C$'s domain as needed, so that the starting constant indexing can be the empty one.}.
% \end{itemize}

\paragraph*{Compilation}
Further down the compilation chain the loop structure is usually partially or completely lost.
We cannot rely on it any more to keep track of the original source code iterations.
We therefore add, alongside the \s{emit} instruction, two other sequential instructions $\sop{ind\_reset}k$ and $\sop{ind\_inc}k$ whose only effect is to reset to 0 (resp.\ increment by 1) the loop index $i_k$. These instructions
will keep track of points in the code corresponding to loop entrances and continuations respectively.

The first step of compilation from $\iota\ell\imp$ consists of prefixing the translation of an indexed loop $\lbl{i_k}:\s{while}\ b\ \s{do}\ S$ with $\sop{ind\_reset}k$ and postfixing the translation of its body $S$ with $\sop{ind\_inc}k$.
Later in the compilation chain we must propagate the instructions dealing with cost labels.

We would like to stress the fact that this machinery is only needed to give a suitable semantics of observables on which preservation proofs can be done.
By no means are the added instructions and the constant indexing in the state meant to change the actual (let us say denotational) semantics of the programs.
In this regard the two new instructions have a similar role as the \s{emit} one.
A forgetful mapping of everything (syntax, states, operational semantics rules) can be defined erasing all occurrences of cost labels and loop indexes, and the result will always be a regular version of the language considered.

\paragraph*{Stating the preservation of semantics}
In fact, the statement of preservation of semantics does not change at all, if not for considering traces of evaluated indexed cost labels rather than traces of plain ones. So every pass will still need to enjoy
\hyperref[eq:preservation]{property~(\ref*{eq:preservation})}.

\subsection{Dependent costs in the source code}
\label{ssec:depcosts}\label{ssec4}
The task of producing dependent costs from constant costs induced by indexed labels is quite technical.
Before presenting it here, we would like to point out that the annotations produced by the procedure described in this subsection, even if correct, can be enormous and unreadable. The prototype compiler employs simplifications that will not be documented here to mitigate this problem.

Upon compiling the indexed labelling $\Ell^\iota(P)$ of an
\imp{} program $P$, we may still apply the machinery described in~\cite{labeling}
and sketched in \autoref{sec:labelling} and get a statically computed cost
mapping from \emph{indexed} labels to naturals.

As we need to annotate the source code, we want a way to express and
compute the costs of cost labels. In order to do so, we have to group
the costs of single indexed labels with the same atom.
In order to do so we introduce \emph{dependent costs}.

Let us suppose that for the sole purpose of annotation, we have available in the language C-like conditional ternary expressions of the form $\tern e {f_1}{f_2},$
and that we have access to common operators on integers such as equality, order and modulus.

\paragraph*{Simple conditions}

First, we need to shift from \emph{transformations} of loop indexes to \emph{conditions} on them.
We identify a set of conditions on natural numbers which are able to express the image of any composition of simple expressions.
\emph{Simple conditions} are of three possible forms:
$$ p \gramm  {i_k=n} | {i_k\ge n} | {i_k\bmod a = b\wedge i_k\ge n} $$

Given a simple condition $p$ and a constant indexing $C$ we can easily define when $p$ holds for $C$ (written $p|_C$):
it suffices to substitute the formal indexes with their value in $C$.
A \emph{dependent cost expression} is an expression built solely out of integer constants and ternary expressions with simple conditions at their head, \ie
$K\gramm n \mid \tern p {K_1} {K_2}.$
Given a dependent cost expression $K$ where all of the loop indexes appearing in it are in the domain of a constant indexing $C$, we can easily define the value $K|_C\in \mathbb N$ by evaluating the heads of all
ternary expressions in $C$.
% $$n|_C\ass n,\qquad (\tern p K_1 K_2)|_C\ass
% \begin{cases}
%   K_1|_C& \text{if $p|_C$,}\\
%   K_2|_C& \text{otherwise.}
% \end{cases}$$

Every simple expression $e$ corresponds to a simple condition $p(e)$ which expresses the set of values that $e$ can take.
Following is the definition of such a relation%
\footnote{We recall that in this development, loop indexes are always mapped to simple expressions over the same index.
If it was not the case, the condition obtained from an expression should be on the mapped index, not the indeterminate of the simple expression.
We leave all generalisations of what we present here for further work}:
$$
\begin{gathered}
p(0*i_k+b)\ass ({i_k = b})
\qquad\qquad
p(1*i_k+b)\ass ({i_k \ge b})\\
p(a*i_k+b)\ass ({i_k\bmod a = b' \wedge i_k \ge b)}\quad\text{if $a>1$, where $b' = b\bmod a$}.
\end{gathered}
$$
The fact that this mapping has sense is stated by the following fact.
\begin{fact}
\label{fact:simple_conditions_ok}
For every expression $e$ on $i_k$, $p(e)|_{(i_k\mapsto c)}$ iff there is a
constant $d$ such that $e|_{(i_k\mapsto d)}= c$.
\end{fact}

\paragraph*{From indexed costs to dependent ones}
Suppose we are given a mapping $\kappa$ from indexed labels to natural numbers.
We must transform it to a mapping (identified, by abuse of notation, with the same symbol $\kappa$) from atoms to dependent expressions.
The reader uninterested in the technical details explained below can get a grasp of how this is done by going through the example
in \autoref{ssec:detailedex}.

We will allow indexings to start from other index variables than $i_0$.
Let $\mathbb S$ be the set of sets of indexings with fixed domain.
Formally:
$$\mathbb S\ass \setof{S \mid S\subseteq \setof{i_h\mapsto e_h,\dots,i_k\mapsto e_k}
\text{for some $h\le k$ and $e_i$'s}},$$
For every set $S\in \mathbb S$, we are in one of the following three mutually exclusive cases:
\begin{itemize}
\item
$S=\emptyset$.
\item
$S=\{\varepsilon\}$, \ie a singleton of the empty indexing.
\item
There is $i_h\mapsto e$ such that $S$ can be decomposed in $(i_h\mapsto e)S'+S''$, with $S'\neq \emptyset$ and none of the sequences in $S''$ start with $e$.
Here $(i_h\mapsto e)S'$ denotes prepending $i_h\mapsto e$ to all elements of $S'$, while $+$ is disjoint union.
\end{itemize}

The above classification can serve as the basis of a definition by recursion on $n+\card S$ where $n$ is the size of indexings in $S$ and $\card S$ is its cardinality.
Indeed in the third case in $S'$ the size of indexings decreases strictly (and cardinality does not increase) while for $S''$ the size of tuples remains the same but cardinality strictly decreases.
The expression $e$ of the third case can be chosen as minimal for some total order\footnote{The specific order used does not change the correctness of the procedure, but different orders can give more or less readable results. An empirically ``good'' order is the lexicographic one, with $a*i_k+b \le a'*i_k+b'$ if $a<a'$ or $a=a'$ and $b\le b'$.}.

We first define the auxiliary function $ \kappa^\alpha_I$, parametrized
by atoms and $0$-based indexings, and going from $\mathbb S$
to dependent expressions, using the previous classification of elements in
$\mathbb S$.
$$
% \begin{aligned}
\kappa^\alpha_L(\emptyset) \ass 0\qquad
\kappa^\alpha_L(\{\varepsilon\}) \ass \kappa(\alpha\la L\ra) \qquad
\kappa^\alpha_L((i_h\mapsto e)S'+S'') \ass \tern{p(e)}{\kappa^\alpha_{L(i_k\mapsto e)}(S')}{\kappa^\alpha_L(S'')}
% \end{aligned}
$$
Finally the wanted dependent cost mapping is defined by
\begin{equation}
\label{eq:dependentcost}
\kappa(\alpha)\ass\kappa^\alpha_\varepsilon(\{\,L\mid \alpha\la L\ra \text{ appears in the compiled code}\,\})
\end{equation}
where one must notice that the set of indexings of an atom appearing in the code inhabits $\mathbb S$ because
the domain of all indexings is fixed by the number of nested loops in the source code.

The correctness of the above formula, which is a consequence of \autoref{fact:simple_conditions_ok},
can be stated as the following.
\begin{fact}
\label{fact:dependent_cost_ok}
If there is no overlap (see \autoref{fact:nonoverlap}), and $\alpha\la I\ra|_C=\alpha\la D\ra$
for $\alpha\la I\ra$ occurring in the compiled code, then
$\kappa(\alpha)|_D=\kappa(\alpha\la I\ra)$.
\end{fact}
The no overlap hypothesis ensures that if we are in the third case
$\kappa^\alpha_L((i_h\mapsto e)S'+S'')$ of the
formula above and $I=L,J$ with $J\in S''$, then $p(e)|_D$ does not hold.

\paragraph*{Indexed instrumentation}
The \emph{indexed instrumentation} generalises the instrumentation as presented
in~\cite{labeling} and sketched in \autoref{sec:labelling}.
We described above how cost atoms can be mapped to dependent costs.
The indexed instrumentation $\mathcal I^\iota$ must also insert code dealing
with loop indexes.
As instrumentation is done on the code produced by the labelling phase, all cost labels are indexed by identity indexings.
The relevant cases of the recursive definition (supposing $c$ is the cost variable) are then:
$$
\begin{aligned}
\mathcal I^\iota(\lbl{\alpha\la Id_k\ra}:S) &= c\ass c + \kappa(\alpha);\mathcal I^\iota(S)\\
\mathcal I^\iota(\lbl i_k : \sop{while}b\sbin{do}S) &=
  i_k \ass 0; \sop{while}b\sbin{do}(\mathcal I^\iota (S); i_k \ass i_k + 1)
\end{aligned}
$$

This means that instrumentation internalises an index state $C$ as the actual values of
variables $i_0,\ldots$, and when a cost must be registered it adds to the global
cost variable the value $\kappa(\alpha)|_C$ using the current index state.

Suppose we guarantee the semantic correctness of the compilation and the
fact that we never produce overlapping indexed labels (\autoref{fact:nonoverlap}
for loop transformations, trivial for other passes).
The correctness of the instrumentation then follows from \autoref{fact:dependent_cost_ok}.
Indeed if the source code emits $\alpha \la C\ra$, by semantic correctness we have
the corresponding point in the execution of the compiled code emitting the same,
which means that we have encountered $\alpha\la I\ra$ under index state
$D$ such that $\alpha\la I\ra|_D=\alpha\la C\ra$. Moreover the index state in
the labelled source is $C$, as all indexings are identities. It follows that
when evaluating the instrumentation $c \ass c + \kappa(\alpha)$, we
add to the cost variable the amount $\kappa(\alpha)|_C=\kappa(\alpha\la I\ra)$,
which is correct if the static analysis correctly analysed the cost.

\subsection{A detailed example}\label{ssec:detailedex}
Take the program in \autoref{fig:example1}. Its initial labelling is
shown in \autoref{subfig:example1labeled}.
Supposing for example, $n=3$
the trace of the program will be
$$\alpha\la\ra\,\beta\la 0 \ra\, \delta\la 0\ra\,\beta\la 1\ra\,\gamma\la 1,0\ra\,
\delta\la 1\ra\,\beta\la 2\ra\,\gamma\la 2,0\ra\,\gamma\la 2, 1\ra\,\delta\la 2\ra\,
\epsilon\la\ra$$
Now let us apply the transformations of \autoref{fig:example1} with the additional
information detailed in \autoref{fig:indexed_loop_transformations}. The result is shown in
\autoref{subfig:example1transformed}. One can check that the transformed code leaves the same trace when executed.
\begin{figure}[!ht]
{}\hfill
\begin{minipage}{4.4cm}
\begin{tikzpicture}
\begin{lstnode}{codelabld}{4.15cm}
$\lbl{\alpha\la \ra}:s\ass 0$;
$i\ass 0$;
$\lbl{i_0}:$ while $i<n$ do
    $\lbl{\beta\la i_0\ra}: p\ass 1$;
    $j\ass 1$;
    $\lbl{i_1}:$ while $j \le i$ do
        $\lbl{\gamma\la i_0, i_1\ra} : p\ass j*p$;
        $j\ass j+1$;
    $\lbl{\delta\la i_0\ra} : s\ass s+p$;
    $i\ass i+1$;
$\lbl{\epsilon\la\ra}:$ skip
\end{lstnode}
\end{tikzpicture}
\subcaption{}
\label{subfig:example1labeled}
\end{minipage}
\hfill
\begin{minipage}{9.25cm}
\begin{tikzpicture}
\begin{lstnode}{code3}{9cm}
$\lbl{\alpha\la\ra}:s\ass 0$;
$i\ass 0$;
if $i<n$ then
    $\lbl{\beta\la0\ra}:p\ass 1$;
    $j\ass 1$;
    $\lbl{i_1}:$ while $j\le i$ do
        $\lbl{\gamma\la0,i_1\ra}:p\ass j*p$;
        $j\ass j+1$;
    $\lbl{\delta\la 0\ra}:s\ass s+p$;
    $i\ass i+1$;
    $\lbl{i_0}:$ while $i<n$ do
        $\lbl{\beta\la 2*i_0+1\ra}:p\ass 1$;
        $j\ass 1$;
        if $j\le i$ then
            $\lbl{\gamma\la 2*i_0+1,0\ra}:p\ass j*p$;
            $j\ass j+1$;
            if $j\le i$ then
                $\lbl{\gamma\la 2*i_0+1,1\ra}:p\ass j*p$;
                $j\ass j+1$;
                $\lbl{i_1}:$ while $j\le i$ do
                    $\lbl{\gamma\la 2*i_0+1,2*i_1+2\ra}:p\ass j*p$;
                    $j\ass j+1$;
                    if $j\le i$ then
                        $\lbl{\gamma\la 2*i_0+1,2*i_1+3\ra}:p\ass j*p$;
                        $j\ass j+1$;
        $\lbl{\delta\la 2*i_0+1\ra}:s\ass s+p$;
        $i\ass i+1$;
        if $i<n$ then
            $\lbl{\beta\la 2*i_0+2\ra}:p\ass 1$;
            $j\ass 1$;
            $\lbl{i_1}:$ while $j\le i$ do
                $\lbl{\gamma\la 2*i_0+2,2*i_1\ra}:p\ass j*p$;
                $j\ass j+1$;
                if $j\le i$ do
                    $\lbl{\gamma\la 2*i_0+2,2*i_1+1\ra}:p\ass j*p$;
                    $j\ass j+1$;
            $\lbl{\delta\la 2*i_0+2\ra}:s\ass s+p$;
            $i\ass i+1$;
$\lbl{\epsilon\la \ra}:{}$skip
\end{lstnode}
\end{tikzpicture}
\subcaption{}
\label{subfig:example1transformed}
\end{minipage}
\hfill{}
\caption{The result of indexed labeling and reindexing loop transformations on the
program in \autoref{fig:example1}. A single \s{skip} after the $\delta$ label has
been suppressed, and we are writing $\alpha\la e_0, \ldots, e_k\ra$ for
$\alpha\la i_0\mapsto e_0, \ldots, i_k\mapsto e_k\ra$.}
\end{figure}

\begin{figure}
\centering\small
\newcommand{\indetern}[4][b]{\begin{tabular}[#1]{nLn}(#2)\mathrel ?{}\\\hskip 10pt #3:{}\\\hskip 10pt #4\end{tabular}}
\def\tern#1#2#3{(#1)\mathrel ? #2:#3}
{}\hfil
\begin{subfigure}{75.7pt}
\centering
$\begin{aligned}
  &0, i_1\\
  &2*i_0+1, 0\\
  &2*i_0+1, 1\\
  &2*i_0+1, 2*i_1+2\\
  &2*i_0+1, 2*i_1+3\\
  &2*i_0+2, 2*i_1\\
  &2*i_0+2, 2*i_1+1
\end{aligned}$
% \hfill$|$
\caption{The indexings of $\gamma$ in \autoref{subfig:example1transformed}.}
\label{subfig:indexings}
\end{subfigure}
\hfil
\begin{subfigure}{176.7pt}
\centering
$
% \kappa^\iota(\gamma)=
\indetern[t]{i_0 = 0}
  {\tern{i_1\ge 0}a0}
  {\indetern{i_0\bmod 2 = 1 \wedge i_0\ge 1}
    {\indetern{i_1=0}
      b
      {\indetern{i_1 = 1}
        c
        {\indetern{i_1\bmod 2 = 0 \wedge i_1\ge 2}
          d
          {\tern{i_1\bmod 2 = 1 \wedge i_1\ge 3}e0}
        }
      }
    }
    {\indetern{i_0\bmod 2 = 0 \wedge i_0\ge 2}
      {\indetern{i_1 \bmod 2 = 0 \wedge i_1 \ge 0}
        f
        {\tern{i_1 \bmod 2 = 1 \wedge i_1 \ge 1}g0}
      }
      0
    }
  }
$
% \hfill$|$
\caption{The dependent cost of $\gamma$ as given by
\hyperref[eq:dependentcost]{equation (\ref*{eq:dependentcost})}.}
\label{subfig:gammadependent}
\end{subfigure}
\hfil
\begin{subfigure}{133.6pt}
$
% \kappa^\iota(\gamma)=
\indetern[t]{i_0 = 0}
  a
  {\indetern{i_0\bmod 2 = 1}
    {\indetern{i_1=0}
      b
      {\indetern{i_1 = 1}
        c
        {\tern{i_1\bmod 2 = 0}
          de
        }
      }
    }
    {\tern{i_1 \bmod 2 = 0}
      fg
    }
  }
$
\caption{The dependent cost of $\gamma$ as simplified by a procedure
not described in this work but implemented in CerCo's compiler.
Further simplifications would be possible if any of the constants turn
out to be equal.}
\label{subfig:gammadependentsimplified}
\end{subfigure}
\hfil{}
\caption{The dependent cost of $\gamma$ in the program
of \autoref{fig:example1}, as transformed in \autoref{subfig:example1transformed}.}
\end{figure}

Let us compute the dependent cost of $\gamma$, supposing no other loop transformations
are done. Ordering its indexings we
have the list in \autoref{subfig:indexings}.
If we denote with $a,b,\ldots,g$ the integer costs statically computed from
the compiled code for each of the indexed occurrences of $\gamma$ in the compiled code in \autoref{subfig:example1transformed},
we obtain, using \hyperref[eq:dependentcost]{equation (\ref*{eq:dependentcost})}
and the order of indexings in \autoref{subfig:indexings}, the depedent cost in
\autoref{subfig:gammadependent}. Applying some simplifications that are not
documented here but that are implemented in CerCo's untrusted prototype,
we obtain the equivalent dependent cost in \autoref{subfig:gammadependentsimplified}.

One should keep in mind that the example was wilfully complicated, in practice
the cost expressions produced have rarely more clauses
than the number of nested loops containing the annotation.

\section{Future work}\label{sec:future}
For the time being, indexed labels are only implemented in the untrusted Ocaml compiler, while they are not present yet in the code on which the computer assisted proof can be carried out (in case of CerCo's
project, the tool used is Matita~\cite{matita}). Porting them should pose no significant problem.
Once ported, the task of proving properties about them in Matita can begin.

Because most of the executable operational semantics of the languages across the front end and the back end are oblivious to cost labels, it should be expected that the bulk of the semantic preservation proofs that still needs to be done will not get any harder because of indexed labels.
The only trickier point that we foresee would be in the translation of \s{Clight} to \s{Cminor} (the first pass of CerCo's compiler's front-end), where we pass from structured indexed loops to atomic instructions on loop indexes.

An invariant which should probably be proved and provably preserved along the compilation chain is the non-overlap of indexings for the same atom.
Then, supposing cost correctness for the unindexed approach, the indexed one will just need to amend the proof by stating
$$\forall C\text{ constant indexing}.\forall \alpha\la I\ra\text{ appearing in the compiled code}.
  \kappa(\alpha)|_{I_C} = \kappa(\alpha\la I \ra).
$$
Here, $C$ represents a snapshot of loop indexes in the compiled code, while $I\circ C$ is the corresponding snapshot in the source code.
Semantics preservation will ensure that when, with snapshot $C$, we emit $\alpha\la I\ra$ (that is, we have $\alpha\la I\circ C\ra$ in the trace), $\alpha$ must also be emitted in the source code with indexing $I\circ C$, so the cost $\kappa(\alpha)\circ (I\circ C)$ applies.

Aside from carrying over the proofs, we would like to extend the approach to more loop transformations.
Important examples are loop inversion (where a for loop is reversed, usually to make iterations appear to be truly independent) or loop interchange (where two nested loops are swapped, usually to have more loop invariants or to enhance strength reduction).
This introduces interesting changes to the approach, where we would have indexings such as:
$$i_0\mapsto n - i_0\quad\text{or}\quad i_0\mapsto i_1, i_1\mapsto i_0.$$
In particular dependency over actual variables of the code would enter the frame, as indexings would depend on the number of iterations of a well-behaving guarded loop (the $n$ in the first example).

Finally, as stated in the introduction, the approach should allow some integration of techniques for cache analysis, a possibility that for now has been put aside as the standard 8051 target architecture for the CerCo project lacks a cache.
Two possible developments for this line of work present themselves:
\begin{enumerate}
\item
One could extend the development to some 8051 variants, of which some have been produced with a cache.
\item
One could make the compiler implement its own cache: this cannot apply to \textsc{ram} accesses of the standard 8051 architecture, as the difference in cost of accessing the two types of \textsc{ram} is only one clock cycle, which makes any implementation of cache counterproductive.
So for this proposal, we could either artificially change the accessing cost of \textsc{ram} of the model just for the sake of possible future adaptations to other architectures, or otherwise model access to an external memory by means of the serial port of the microcontroller.
\end{enumerate}

% 
% \newpage
% 
% \includepdf[pages={-}]{plugin.pdf}
% 
% 
% \newpage
% 
% \includepdf[pages={-}]{fopara.pdf}
% 
% 
% \newpage
% 
% \includepdf[pages={-}]{tlca.pdf}
% 
\bibliographystyle{eptcs}
\bibliography{inc/bib}

\newcommand{\online}[1]{Available at \url{#1}}
\begin{thebibliography}{10}
\providecommand{\bibitemdeclare}[2]{}
\providecommand{\surnamestart}{}
\providecommand{\surnameend}{}
\providecommand{\urlprefix}{Available at }
\providecommand{\url}[1]{\texttt{#1}}
\providecommand{\href}[2]{\texttt{#2}}
\providecommand{\urlalt}[2]{\href{#1}{#2}}
\providecommand{\doi}[1]{doi:\urlalt{http://dx.doi.org/#1}{#1}}
\providecommand{\bibinfo}[2]{#2}

\bibitemdeclare{misc}{absint}
\bibitem{absint}
\emph{\bibinfo{title}{{AbsInt} Angewandte Informatik}}.
\newblock \urlprefix\url{http://www.absint.com/}.

\bibitemdeclare{misc}{cerco}
\bibitem{cerco}
\emph{\bibinfo{title}{Certified Complexity ({CerCo}), {FET}-Open {EU}
  Project}}.
\newblock \urlprefix\url{http://cerco.cs.unibo.it/}.

\bibitemdeclare{misc}{framac}
\bibitem{framac}
\emph{\bibinfo{title}{Frama-C software analyzers}}.
\newblock \urlprefix\url{http://frama-c.com/}.

\bibitemdeclare{misc}{matita}
\bibitem{matita}
\emph{\bibinfo{title}{Matita}}.
\newblock \urlprefix\url{http://matita.cs.unibo.it/}.

\bibitemdeclare{unpublished}{D2.2}
\bibitem{D2.2}
\bibinfo{author}{Roberto~M. \surnamestart Amadio\surnameend},
  \bibinfo{author}{Nicolas \surnamestart Ayache\surnameend},
  \bibinfo{author}{Yann \surnamestart R{\'e}gis-Gianas\surnameend} \&
  \bibinfo{author}{Ronan \surnamestart Saillard\surnameend}
  (\bibinfo{year}{2010}): \emph{\bibinfo{title}{Prototype implementation}}.
\newblock \bibinfo{note}{Deliverable 2.2 of Project FP7-ICT-2009-C-243881
  CerCo. \online{http://cerco.cs.unibo.it/}}.

\bibitemdeclare{inproceedings}{labeling}
\bibitem{labeling}
\bibinfo{author}{Nicholas \surnamestart Ayache\surnameend},
  \bibinfo{author}{Roberto~M. \surnamestart Amadio\surnameend} \&
  \bibinfo{author}{Yann \surnamestart R{\'e}gis-Gianas\surnameend}
  (\bibinfo{year}{2012}): \emph{\bibinfo{title}{Certifying and Reasoning on
  Cost Annotations in C Programs}}.
\newblock In \bibinfo{editor}{Mari{\"e}lle \surnamestart Stoelinga\surnameend}
  \& \bibinfo{editor}{Ralf \surnamestart Pinger\surnameend}, editors: {\sl
  \bibinfo{booktitle}{FMICS}}, {\sl \bibinfo{series}{Lecture Notes in Computer
  Science}} \bibinfo{volume}{7437}, \bibinfo{publisher}{Springer}, pp.
  \bibinfo{pages}{32--46}, \doi{10.1007/978-3-642-32469-7\_3}.

\bibitemdeclare{article}{cacheprediction}
\bibitem{cacheprediction}
\bibinfo{author}{Christian \surnamestart Ferdinand\surnameend} \&
  \bibinfo{author}{Reinhard \surnamestart Wilhelm\surnameend}
  (\bibinfo{year}{1999}): \emph{\bibinfo{title}{Efficient and Precise Cache
  Behavior Prediction for Real-TimeSystems}}.
\newblock {\sl \bibinfo{journal}{Real-Time Syst.}} \bibinfo{volume}{17}, pp.
  \bibinfo{pages}{131--181}, \doi{10.1023/A:1008186323068}.

\bibitemdeclare{}{scade}
\bibitem{scade}
\bibinfo{author}{Xavier \surnamestart Fornari\surnameend}:
  \emph{\bibinfo{title}{Understanding how {SCADE} suite {KCG} generates safe
  {C} code}}.
\newblock \bibinfo{note}{White paper, Esterel Technologies.
  \online{http://www.esterel-technologies.com/technology/WhitePapers/}}.

\bibitemdeclare{article}{CompCert}
\bibitem{CompCert}
\bibinfo{author}{Xavier \surnamestart Leroy\surnameend} (\bibinfo{year}{2009}):
  \emph{\bibinfo{title}{Formal verification of a realistic compiler}}.
\newblock {\sl \bibinfo{journal}{Commun. ACM}}
  \bibinfo{volume}{52}(\bibinfo{number}{7}), pp. \bibinfo{pages}{107--115},
  \doi{10.1145/1538788.1538814}.

\bibitemdeclare{article}{PRE}
\bibitem{PRE}
\bibinfo{author}{E.~\surnamestart Morel\surnameend} \&
  \bibinfo{author}{C.~\surnamestart Renvoise\surnameend}
  (\bibinfo{year}{1979}): \emph{\bibinfo{title}{Global optimization by
  suppression of partial redundancies}}.
\newblock {\sl \bibinfo{journal}{Commun. ACM}} \bibinfo{volume}{22}, pp.
  \bibinfo{pages}{96--103}, \doi{10.1145/359060.359069}.

\bibitemdeclare{book}{morgan}
\bibitem{morgan}
\bibinfo{author}{Robert \surnamestart Morgan\surnameend}
  (\bibinfo{year}{1998}): \emph{\bibinfo{title}{Building an Optimizing
  Compiler}}.
\newblock \bibinfo{publisher}{Digital Press}.

\bibitemdeclare{book}{muchnick}
\bibitem{muchnick}
\bibinfo{author}{Steven~S. \surnamestart Muchnick\surnameend}
  (\bibinfo{year}{1997}): \emph{\bibinfo{title}{Advanced Compiler Design and
  Implementation}}.
\newblock \bibinfo{publisher}{Morgan Kaufmann}.

\bibitemdeclare{article}{WCETsurvey}
\bibitem{WCETsurvey}
\bibinfo{author}{Reinhard \surnamestart Wilhelm\surnameend},
  \bibinfo{author}{Jakob \surnamestart Engblom\surnameend},
  \bibinfo{author}{Andreas \surnamestart Ermedahl\surnameend},
  \bibinfo{author}{Niklas \surnamestart Holsti\surnameend},
  \bibinfo{author}{Stephan \surnamestart Thesing\surnameend},
  \bibinfo{author}{David~B. \surnamestart Whalley\surnameend},
  \bibinfo{author}{Guillem \surnamestart Bernat\surnameend},
  \bibinfo{author}{Christian \surnamestart Ferdinand\surnameend},
  \bibinfo{author}{Reinhold \surnamestart Heckmann\surnameend},
  \bibinfo{author}{Tulika \surnamestart Mitra\surnameend},
  \bibinfo{author}{Frank \surnamestart Mueller\surnameend},
  \bibinfo{author}{Isabelle \surnamestart Puaut\surnameend},
  \bibinfo{author}{Peter~P. \surnamestart Puschner\surnameend},
  \bibinfo{author}{Jan \surnamestart Staschulat\surnameend} \&
  \bibinfo{author}{Per \surnamestart Stenstr{\"o}m\surnameend}
  (\bibinfo{year}{2008}): \emph{\bibinfo{title}{The worst-case execution-time
  problem - overview of methods and survey of tools}}.
\newblock {\sl \bibinfo{journal}{ACM Trans. Embedded Comput. Syst.}}
  \bibinfo{volume}{7}(\bibinfo{number}{3}), \doi{10.1145/1347375.1347389}.

\bibitemdeclare{inproceedings}{findingbugs}
\bibitem{findingbugs}
\bibinfo{author}{Xuejun \surnamestart Yang\surnameend}, \bibinfo{author}{Yang
  \surnamestart Chen\surnameend}, \bibinfo{author}{Eric \surnamestart
  Eide\surnameend} \& \bibinfo{author}{John \surnamestart Regehr\surnameend}
  (\bibinfo{year}{2011}): \emph{\bibinfo{title}{Finding and understanding bugs
  in C compilers}}.
\newblock In \bibinfo{editor}{Mary~W. \surnamestart Hall\surnameend} \&
  \bibinfo{editor}{David~A. \surnamestart Padua\surnameend}, editors: {\sl
  \bibinfo{booktitle}{PLDI}}, \bibinfo{publisher}{ACM}, pp.
  \bibinfo{pages}{283--294}, \doi{10.1145/1993498.1993532}.

\end{thebibliography}

\end{document}